In Silico Prediction of Blood-Brain Barrier Permeability of Chemical Compounds through Molecular Feature Modeling

Authors: T. Jain (1), P. Kumar (2)
        ((1) Thomas Jefferson High School for Science and Technology, (2) Independent
        Researcher)

The introduction of computational techniques to analyze chemical data has given rise to the analytical study of biological systems, known as "bioinformatics". One facet of bioinformatics is using machine learning (ML) technology to detect multivariable trends in various cases. Amongst the most pressing cases is predicting blood-brain barrier (BBB) permeability. The development of new drugs to treat central nervous system disorders presents unique challenges due to poor penetration efficacy across the blood-brain barrier. In this research, we aim to mitigate this problem through an ML model that analyzes chemical features. To do so: (i) An overview into the relevant biological systems and processes as well as the use case is given. (ii) Second, an in-depth literature review of existing computational techniques for detecting BBB permeability is undertaken. From there, an aspect unexplored across current techniques is identified and a solution is proposed. (iii) Lastly, a two-part *in silico* model to quantify likelihood of permeability of drugs with defined features across the BBB through passive diffusion is developed, tested, and reflected on. Testing and validation with the dataset determined the predictive logBB model's mean squared error to be around 0.112 units and the neuroinflammation model's mean squared error to be approximately 0.3 units, outperforming all relevant studies found.

**Introduction**

**Background**
Drug development is a lengthy, complex, and costly process, entrenched with a high degree of risk surrounding a drug's success. The development of a single prescription medicine that is approved for marketing is estimated to cost drugmakers $2.6 billion. The challenge is amplified for the development of central nervous system (CNS) drugs. CNS drugs typically take 20% longer to develop and 38% longer to approve than non-CNS drugs, with a failure rate of 85% (1). The main reason behind the rejection rate is the poor penetration efficacy across the blood-brain barrier (BBB), limiting the growth of the neurotherapeutics field (2, 3, 4). The BBB comprises epithelial-like tight junctions that limit diffusion from the blood to the brainspace to molecules with a weight <400 Da and hydrogen bond count <8 bonds; the tight junctions act as a physical and biochemical barrier between the CNS and the bloodstream, maintaining the vital homeostasis of the brain (5, 6). The BBB shields the brain from infectious and toxic substances, but also restricts the ability of drugs to target specific locations in the brain to treat neurological disorders.

The importance of the BBB for the treatment of longstanding diseases such as Alzheimer's and Parkinson's means its research necessitates more attention in both academic neuroscience and industry programs (3, 7). The highly selective nature of the BBB deters most effort by researchers in developing solutions to neurological diseases and disorders; therefore, the neurotherapeutics front faces a dilemma in that there is a slim number of treatments for the majority of CNS disorders. Without a method to gauge permeability before approaching clinical trials, decades could pass before the treatment of neurological diseases, leading many to live with brain dysfunctionality.

**Goals**
The goal of this project is to create an applicable machine learning model (ML) to predict the permeation value of chemical compounds across the blood-brain barrier and serve as a preliminary step to clinical trials for drug discovery. This includes the design and development of an in-silico model to input a compound's molecular features to calculate logBB values which is a proven metric of permeability across the blood-brain barrier. The preceding step would apply for healthy barriers. The applicability is to be derived from the neuroinflammation model, an unprecedented component which will factor a patient's C-reactive protein (CRP) level and adjust the predicted logBB value for neuroinflammation. This promotes analysis under diseased barriers, serving the intended purpose of creating neurotherapeutic drugs.

**Literature Review**
The BBB holds the most prominent effects in determining pharmacokinetic properties of bioactive drugs in the brain. It is composed of endothelial cells, pericytes, and astrocytes in direct contact with brain tissue and differs from the typical blood vessel because the endothelial cells form tight junctions, heightening the selectivity (4, 8, 9). This allows it to restrict the time course of a compound's absorption into the extracellular brain space. The measure of a molecule's permeability is governed by logBB value, using the formula:

$$logBB = log(\frac{C_{brain}}{C_{blood}})$$

Certain physicochemical descriptors of drug and drug-like compounds that are indicative of molecule binding capacity have heavy influence in determining whether a molecule can diffuse, actively or passively, over the tight junctions of the BBB (10). These drug properties cannot be linked, however, to BBB diffusion without intricate nonlinear computation methods because there are no present equations that can simulate this correlation. Fortunately, computational techniques such as deep learning technology can provide this multidimensional simulation (11).

The detailed literature review was categorized into three types of models. First, standard computational approaches without artificial intelligence (AI) techniques were explored to determine the need for machine learning implementation. Second, ML models that focused on structural quantification as input data were analyzed. Third, similar ML models that rather solely focused on chemical features or had a joint focus that prioritized chemical features were analyzed.

Conflicting literature surrounding the merit of ML approaches compared to traditional regression for analysis of clinical data has encouraged the research of both techniques for prediction of BBB permeability (12, 13). A prominent feature for traditional analysis is taking 3D structures of molecules and quantifying the data into 1D descriptors, and methods such as VolSurf have been utilized for this purpose. Results in the form of correct logBB classifications have had a wide range of accuracy, from 79% to 90%, supposedly credited to variance in technique (14, 15). Other approaches have varied both the input type and method of output. One study calculated logPS values to represent penetration through the BBB and used the values as a dataset to make predictions based on a drug's logD value, polar surface area, and van der Waals surface area of basic atoms. More unique attempts as such don't do classification accuracy but instead $R^2$ values and have had scores tending to be less than 0.75 (16, 17, 18).

Machine learning is a subfield of AI that relies on a computer's ability to learn patterns on its own. Its application is explored in three methods: supervised, unsupervised, and reinforcement learning (19). In the field of drug discovery, supervised learning has proved to be the most common technique due to the need to verify information and inability to cluster data of numerous variables. Similar to standard computation techniques, machine learning approaches have also explored structural descriptors such as cross-sectional area to predict BBB permeability; such models have achieved accuracies as high as 88% but necessitate a larger dataset relative to standard approaches (4, 20). A more recent approach quantified the structure through its molecular fingerprint and used that as its primary descriptor alongside supplementary chemical features. A molecular fingerprint converts a molecule's structure into a bit string which encodes the structure as a descriptor (21). This approach achieved an improved accuracy of 91.9%, boosting the reputability of ML approaches for classification of drug permeability through the BBB (11).

Furthermore, ML can also be leveraged for a much larger variable count that traditional regression techniques cannot handle. The ability to do so has led scientists to hand-select features believed to play a role in BBB permeability. One comprehensive model attempted logistic regression, linear discriminate analysis, k nearest neighbor, C4.5 decision tree, probabilistic neural network, and support vector machine techniques using their custom dataset but was unable to par computational techniques (22). Generating descriptors has become a common technique for these approaches, and programs such as CODES which organizes molecules from a topological point of view have been leveraged to do so (23). This has also given rise to the use of deep neural networks, a technical mimic of the human brain through the use of nodes, to discover underlying relations in drug data (24, 25). Neural networks are a proven computing technology for identifying hidden patterns in raw data and generalizing nonlinear correlations to go beyond a given dataset (26). An important aspect of neural networks is the adaptable training mechanism. The shifting of weights for different types of data allows a model to account for incomplete datasets and varying importance in the molecules' descriptors (27). This allocates room for model improvement without rewriting source code because inputting new, diverse data can be seamlessly integrated (28). Perhaps one of the strongest models in this literature review came from a multi-core SVM method that used drug side effects and indications as inputs for the prediction. This allowed the model to account for non-passive diffusion and achieve an accuracy of 97% with the limitation of being unable to determine how the drug entered (29). The latter is critical in the field of drug discovery as chemical descriptors such as molecule size can point the inability to permeate to a specific cause, and this is necessary for creating permeable drugs.

Despite accurate models having been developed to relate inputs such as structure or features to an output such as logBB, these forecasters are seldom employed because of assumptions of constants in the brain, limited validation, and in some cases, proprietary nature of the model (30).

To counter this issue, this research explored the causes behind variation of BBB permeability in patients with neurological disorders (31, 32). The most influential variance discovered was inflammation. Measured through a patient's acute phase CRP level, inflammation levels have been shown to have a direct correlation to BBB permeability by prompting leptin resistance across the BBB (33, 34). The protein is in the pentraxin family produced primarily in the liver in response to cytokines interleukin-6 and interleukin-1β, both reactive to the inflammatory cycle and thus outlining the rationale behind using the CRP pathway for incorporation neuroinflammation (35, 36).  Wet lab research has determined a CRP threshold of alternance of 2.5 μg/ml; if higher than this, BBB impairment will factor into expected logBB value of chemical compounds (37). The model incorporates the aforementioned threshold as a basis for whether adjustment is needed to the logBB output.

**Methods**

**Data Acquisition**
This project employed a custom-built, verified dataset of 281 molecules with varying permeability values. The following data was extracted from the public PubChem database using Selenium tools and incorporated into the machine learning model: Molecular Weight, Mass, XLogP, Hydrogen Bond Acceptor Count, Hydrogen Bond Donor Count, Rotatable Bond Count, Monoisotopic Mass, Formal Charge, Topological Polar Surface Area, Atom Count, Isotope Atom Count, Atom Stereocenter Count, Bond Stereocenter Count, Covalently-Bonded Unit Count, Vapor Pressure, and Complexity.

**Model Selection**

The project has a defined two-step process to determine numerical permeability of a compound on a patient-by-patient basis. 3D molecular modeling had been extensively researched over decades through deep learning techniques; therefore, a tabular data analysis approach was favored as it potentially had unexplored avenues (11). With background from Alsenan et. al (25), a multi-layer perceptron regression (MPR) model was initially experimented with for the predictive logBB model. The MPR was proficient in determining patterns of high correlation. The issue that arose was outlier accountancy because the model was indeterminate for molecules that fell outside the concentrated features of the dataset. Since the goal of this project was to predict permeability of compounds that are not already existent, this posed a greater risk due to inability to forecast the descriptors of future drugs in development. Ensemble modeling was employed next using methodology from Plisson and Piggott (20). Bagging methods were used to train a series of weak models and combine to create a stronger model. Boosting methods were also used to sequentially train weak models that way multiple techniques could be used, each building off the last. The issue with ensemble modeling was the accuracy (error level) computed was less than that of existing research, although still stronger than the MPR model.

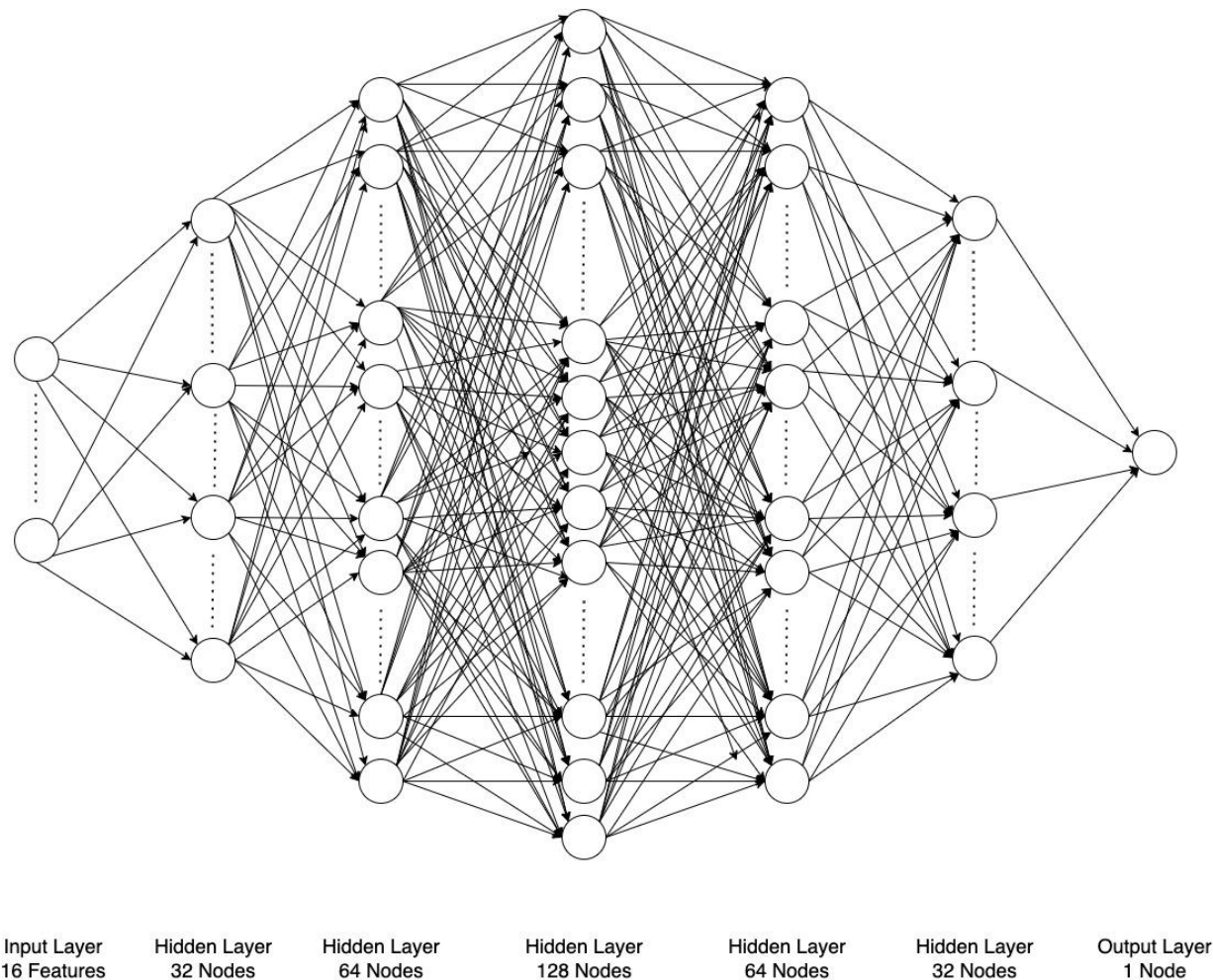

| Input Layer | Hidden Layer | Hidden Layer | Hidden Layer | Hidden Layer | Hidden Layer | Output Layer |
|---|---|---|---|---|---|---|
| 16 Features | 32 Nodes | 64 Nodes | 128 Nodes | 64 Nodes | 32 Nodes | 1 Node |

**Figure 1.** Predictive logBB Model Network Architecture

After experimenting with different model architectures and their performance, a decision was made to go with a fully connected neural network (FCNN). The FCNN, cited as the predictive logBB model, takes in an assortment of preprocessed features. It is a neural network with an input layer of 16 nodes and an output layer with a single node depicting the logBB value. Between the input and the output layer there are 6 hidden layers with a node breakdown per layer of 64, 128, 256, 128, 32, and 1 node(s), respectively, that are trained to learn the logBB value through multivariable analysis. This model was selected on two bases: it was able to learn the significance across the distribution, and it achieved an error level less than both aforementioned models.

The neuroinflammation model works in conjunction with the logBB Model if there is additional information about the C-Reactive Protein levels in a patient. This feature did not have a direct correlation on the logBB values; however, the second and third order feature derivatives were believed to have significant correlations with the logBB values. Hence, the decision was made to use a quadratic polynomial regression model for determining the validity of that correlation.

**Solution**

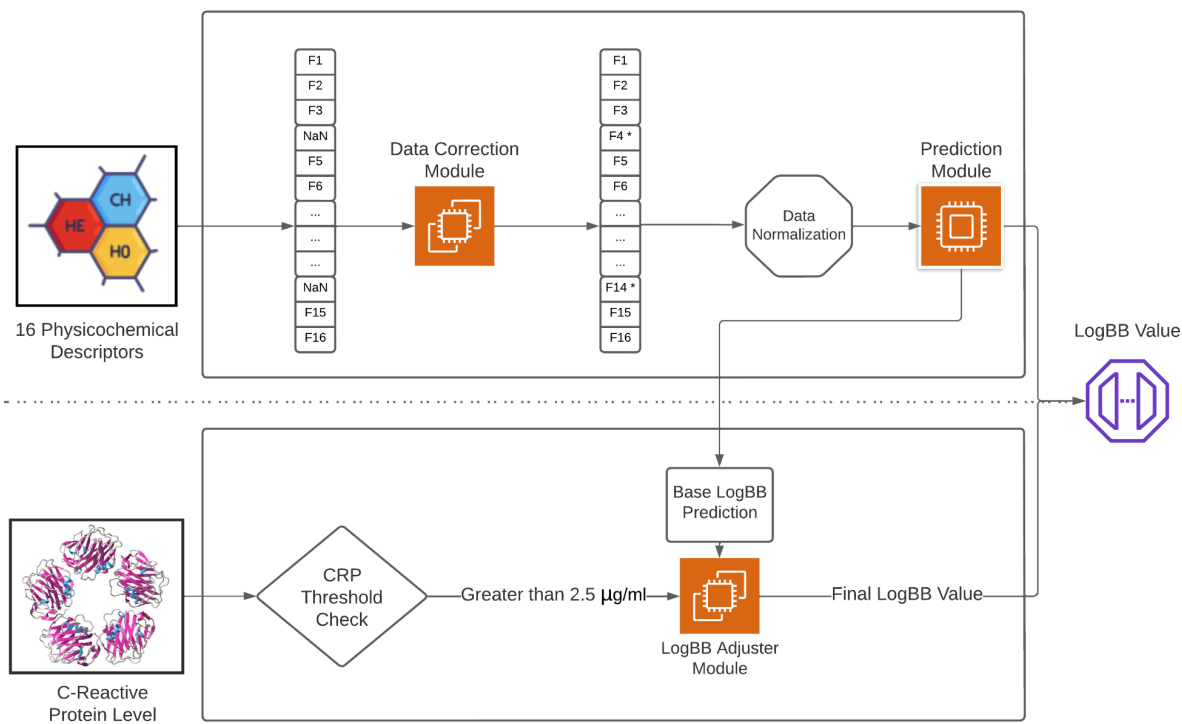

**Figure 2.** Synchronous Model Architecture

The software consists of three key components aside from data acquisition: preprocessing, the predictive logBB model and the neuroinflammation model.

The input receiver takes in the raw input of the 16 aforementioned physicochemical descriptors and, if desired, a patient's CRP level. Preprocessing is split into two consecutive steps: data correction and data normalization. To maximize compound inclusion in the model, missing features from the 16 descriptors were predicted using multiple quadratic polynomial regression models. Yet since data correction was an educated prediction, the model associated lower weights with these values, so the neural network could include a variety of compounds without incorrect or exaggerated permeability output. The complete data undergoes normalization next. The distribution of the training set was evaluated to extract how to bring the mean to zero and the variance to one (univariance). This methodology to normalize is transposed with new drug compounds that enter the model. Following preprocessing, the data is forwarded to the predictive logBB model. It was trained across the dataset derived from PubChem and its error was determined using mean squared error compared to the known logBB value. When the predicted logBB value is generated, this path is halted and the neuroinflammation channel initiates. The CRP value is brought to a threshold checkpoint. If the value is less than 0.0025 g/L, then the inflammation level is deemed insignificant and the original logBB value is the final output. If not, then inflammation can be attributed to a non-healthy BBB and thereby altered restrictiveness. To adjust the drug permeation value, the initial logBB value is sent into the neuroinflammation model, a machine learning quadratic polynomial regression model, along with the CRP level to produce a more accurate logBB value for the drug in trial. The final output layer is a continuous output that predicts the logBB value for the user, also geared to minimize overfitting. The two channel system's output is designed to represent the numerical permeation value through the BBB of a patient with an inflammation level relatively close to the CRP value.

**Training and Validation**
Small batch sizes were used to train the predictive logBB model and to ensure that the network was learning and not memorizing. Small batches also ensured that the network was more efficient and trained faster. Multiple batch sizes were experimented with to observe loss and error for each epoch, and the best one was highlighted. It was observed that accuracy increased linearly with each epoch. It also proved that the network was not overfitting with the given dataset, so applicability was less of a concern. The neuroinflammation model had a more unique process in both training and validation. Since no present research has been made in identifying an equation rather than a correlation, there was no data available to train an in silico model. To acquire data for training the regression model, a common CRP level distribution was pulled from prior research (37). From this, statistical simulations of logBB adjustments were made based on these inputs using the Monte Carlo simulation method. Mean squared error was once again used to predict efficiency of the neuroinflammation model and charts for both models are presented below.

**Results**
The predictive logBB model made 150 passes through the training set and updated the model every 32 sample predictions. Five-fold cross validation was used to summarize the skill of a model and limit unforeseen bias in the dataset. The model error chart and associated hyperparameters are detailed below.

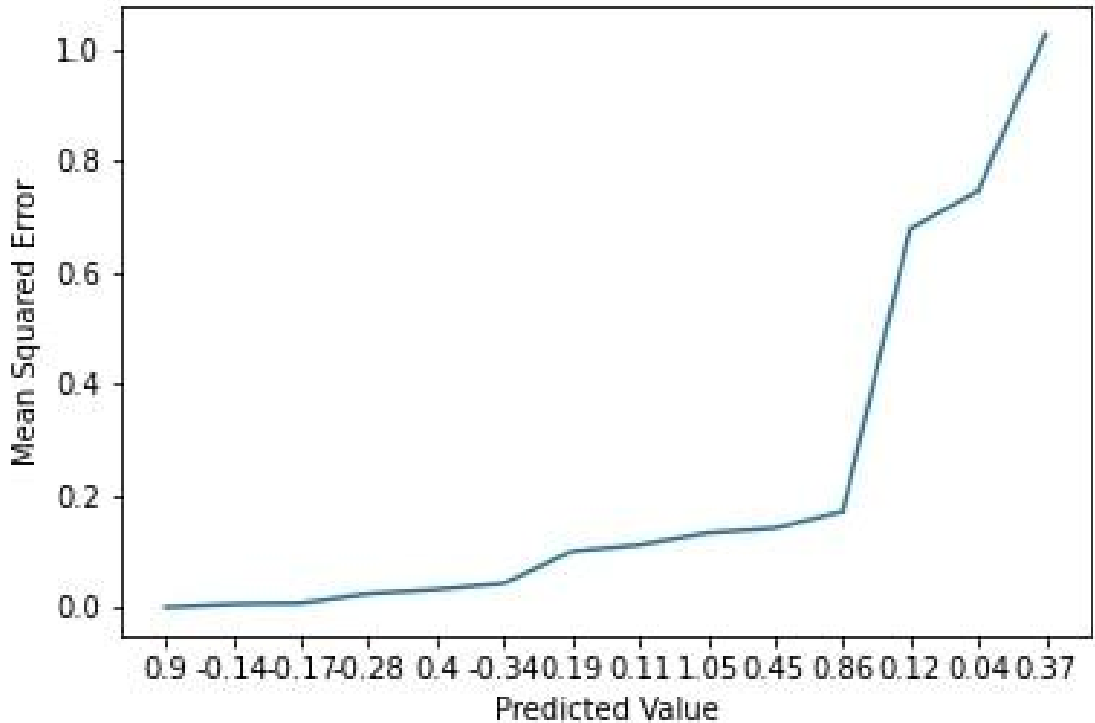

**Figure 3.** Predictive logBB Model Error. Part 1 of 2-Part Model

Predictive logBB Model Hyperparameters:
1. Number of Epochs: 150
2. Batch Size: 32
3. Fold Validation: Five-fold cross validation

The neuroinflammation model countered the lack of defined equations relating CRP levels to logBB values by using machine learning estimators. 50 estimators, or equations, were used to take the CRP distribution and associated effect on the restrictiveness of the BBB to simulate drug permeation and predict logBB values. Since there was no data to compare inflamed logBB values to, error was based on comparison to the same dataset employed in the first model. The purpose of this was to determine if the correlation can be quantified.

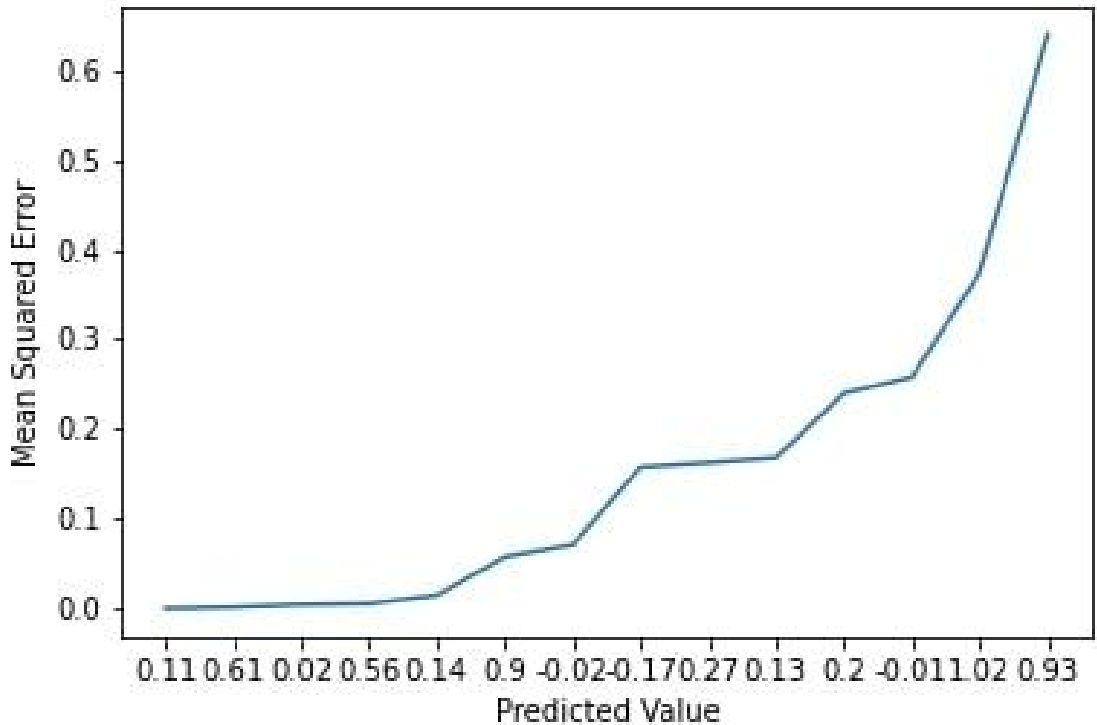

**Figure 4.** Neuroinflammation Model Error. Part 2 of 2-Part Model

Neuroinflammation Model Hyperparameters:
1. Number of Estimators: 50

Our results show that the predictive logBB model achieved a mean squared error of 0.112 and the neuroinflammation model achieved a mean squared error of 0.3. The logBB model surpassed the vast majority of models in modern research, suggesting that the dataset features were appropriate for predicting BBB permeability. The neuroinflammation model, without any prior baseline, has achieved an error comparable to most cited prediction models. This suggests that second and third order derivative correlation of CRP levels to logBB values is mathematically present and quantifiable given a database for drug permeability through inflamed BBBs.

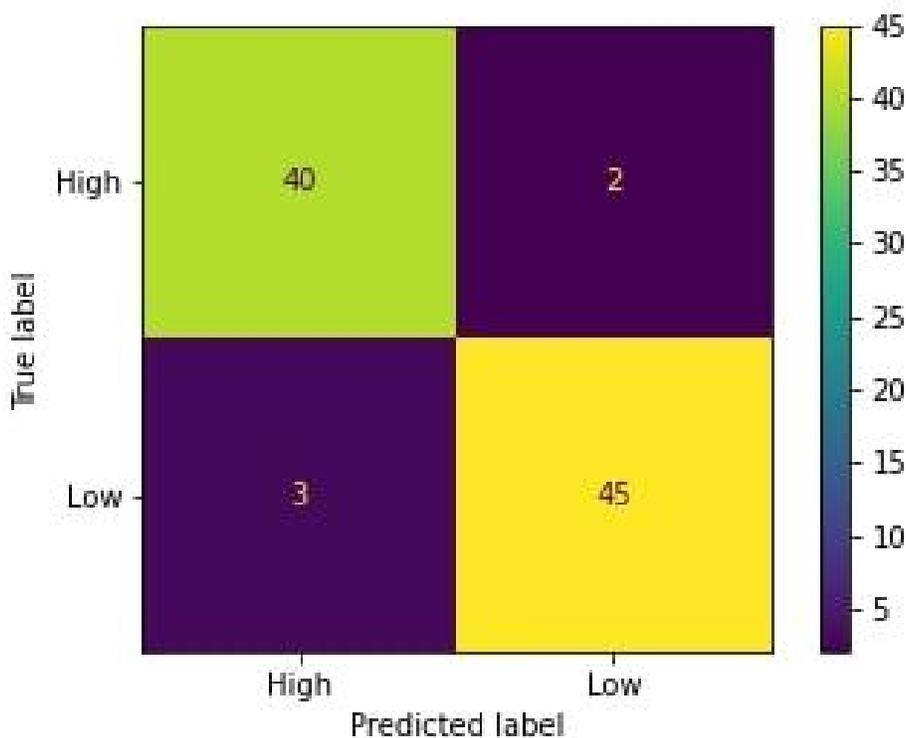

**Figure 5.** Confusion Matrix for Predictive logBB Model

The current model could not be assessed on accuracy because it was not classification-based. To gain further insight into the model's efficacy, a research-backed logBB threshold of 0.3 units was used to categorize predicted values as permeable (>0.3 units) or impermeable (<=0.3 units). From here, the classification could be compared to the true permeation from the dataset. The confusion matrix demonstrates a strong ability to determine if a drug compound can permeate across the BBB. The overarching accuracy of the test set was 94.4% For impermeable compounds, the precision and recall was 95.7% and 93.8%, respectively. For permeable compounds, the precision and recall was 93.0% and 95.2%, respectively. The minimal false positives and false negatives further demonstrates consistency across the range of potential logBB values.

**Discussion**
The model developed in this research surpassed nearly every model that was explored in both the traditional computation and machine learning scope. Those that did perform better had limitations on other fronts such as being unable to point permeability failure to one reason. While the model achieved a notable accuracy for passive diffusion of small molecules, it is important to recognize this model does not address permeation via active diffusion. Another area to be explored would be compiling this model with one that used drug side effects for prediction to include non-passive diffusion predictions. Beyond this, taking away more constants within the BBB would bring models closer to a realistic human brain. More diverse approaches presently serve as the best approach to improving accuracy and applicability.

This project offers a unique add-on to modern day research on BBB permeability of drug compounds. Its design enables targeted clinical trials that takes away one of many assumptions made by pharmaceutical scientists in the drug development process. The ability to gauge within certain inflammation ranges opens a new avenue of drugs that could be introduced for diseases that cause inflammation in these ranges. Rather than an overarching medication that is suited for all patients, various small range neurotherapeutics will mitigate the effects of neurological diseases in a much smaller time period. Paralleled with the current work, this will be a notable stride in the neuroinformatics and drug discovery industries.

The future work of this project will involve organ-on-a-chip biotechnology. Using the simulated blood-brain barrier, the model's applicability can be evaluated in a wet-lab setting. Furthermore, components of this project can be stripped and put into new endeavors in the neuroscience field. One such application is the neuroinflammation readings through CRP levels being used for diagnosis of neurodegenerative diseases. Lastly, the complete model has potential to open a path into precision medicine once explored more through the idea of tailored clinical trials.